\documentstyle[11pt,newpasp,twoside,epsf]{article}
\def\edcomment#1{\iffalse\marginpar{\raggedright\sl#1\/}\else\relax\fi}
\reversemarginpar

\begin{document}
\title{Optimising GAIA: How do we meet the science challenge?}
\author{Gerry Gilmore}
\affil{Institute of astronomy, Madingley Road, Cambridge CB3 0HA, UK}

\begin{abstract}
GAIA has been approved to provide the data needed to quantify
the formation and evolution of the Milky Way Galaxy, and its near
neighbours. That requires study of all four key Galactic stellar
populations: Bulge, Halo, Thick Disk, Thin Disk. The complex analysis
methodologies required to model GAIA kinematic data are being
developed, and in the interim applied to the relatively simple cases
of the satellite dSph galaxies. These methodologies, illustrated here,
show that we will be able to interpret the GAIA data. They also
quantify what data GAIA must provide. 

It is very unlikely in the present design that GAIA will be able to
provide either radial velocities or worthwhile photometry for study of
two of the key science goals: the Galactic Bulge and the (inner)
Galactic old disk. The implication is that the radial velocity
spectrometer and the medium band photometer should be optimised for study
of low density fields -- suitable for their low spatial resolution --
and the broad band photometry must be optimised for inner galaxy
astrophysical  studies.

\end{abstract}

\section{Introduction}

\begin{figure}[h!]
\label{fig1}
\plotfiddle{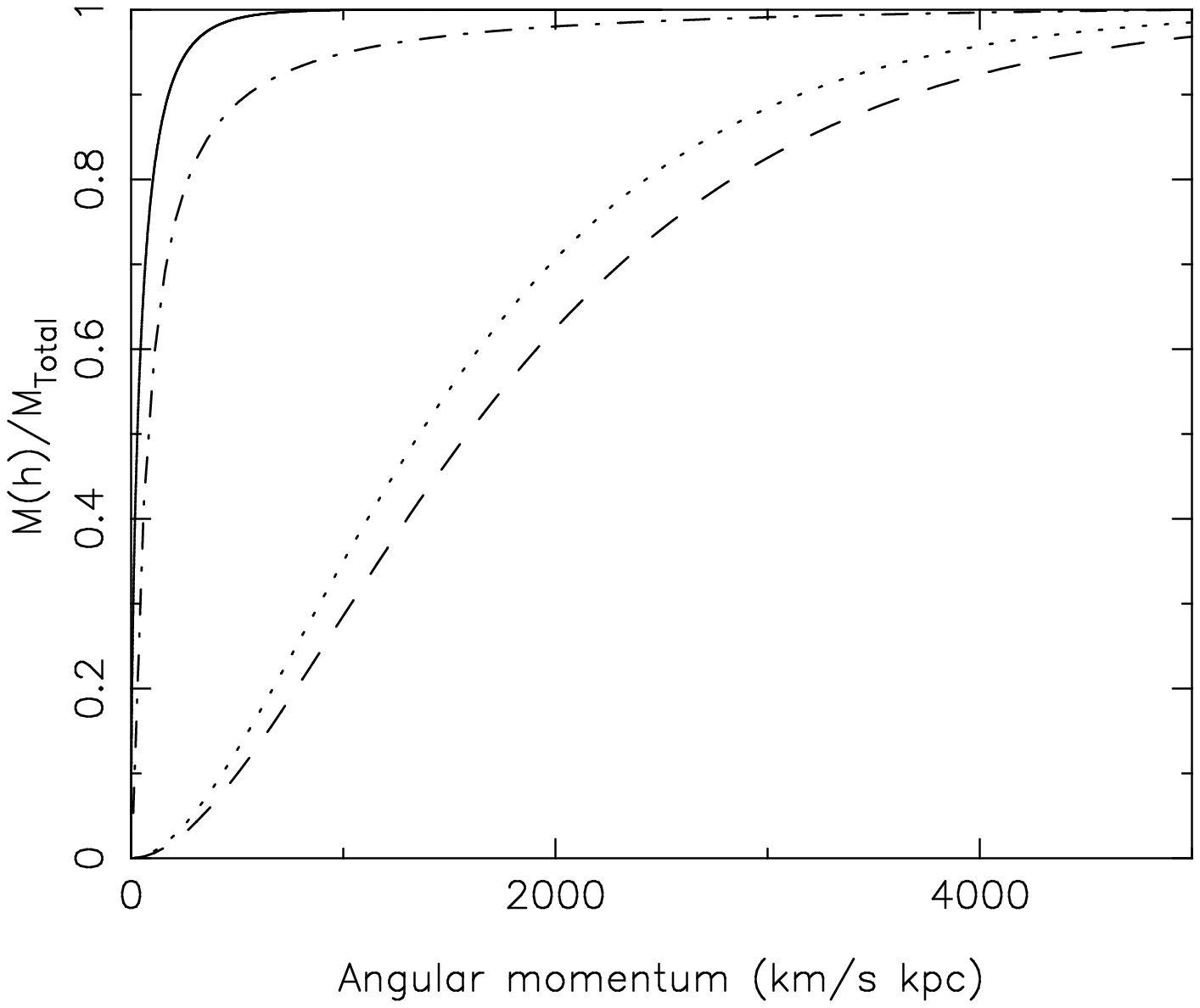}{8cm}{0}{50}{50}{-150}{00}
\caption{Galactic population angular momentum cumulative distribution
functions, showing bulge/halo (solid and dash-dot adjacent curves) and
thin/thick disk (dotted and long-dashed adjacent curves)
dichotomy. This indicates disparate evolutionary histories, and
emphasises the critical need for GAIA to study all four Galactic
components.}
\end{figure}

The GAIA mission has been approved to provide for the first
time a clear picture of the formation, structure, evolution, and
future of the entire Milky Way. In addition, as secondary goals, GAIA
will contribute to many other branches of astrophysics, especially
stellar and solar system minor body astrophysics, with a valuable
contribution to cosmology and fundamental physics.

This clear scientific prioritisation must drive the design, and all
compromises.

Understanding the structure and evolution of the Galaxy requires three
complementary observational approaches: * Carrying out a full census
of all the objects in a large, representative, part of the Galaxy; *
Mapping quantitatively the spatial structure of the Galaxy; *
Measuring the motions of objects in three-dimensions to determine the
gravitational field and the stellar orbits.  In other words, what is
required are complementary measurements of distances (astrometry), 
photometry to determine both extinction and intrinsic stellar
properties, and the radial velocities along our line of sight.

Detailed studies of the Local Group are the key specific test of our
understanding of the formation and growth of structure in the
Universe.  The most significant questions here relevant for GAIA
include the history of most of the stars in the Galaxy: the nature of
the inner Galactic disk and the Galactic Bulge: what are their age,
abundance and assembly histories? Stellar studies locally of course
are of critical general significance: only locally can we determine
the stellar Initial Mass Function directly.  This function directly
controls the chemical and luminosity evolution of the
Universe. Galactic satellite galaxies are proving the most suitable
environs to quantify the nature and distribution of dark matter, and
to test the small scale predictions of hierarchical galaxy formation
models. The Galactic disk itself, of course, is a key test of angular
momentum distributions (figure~1), chemical evolution, and merger
histories, and must eventually provide the most robust information on
Cold Dark Matter.
Such issues as the number of local dwarf
galaxies and the inner CDM profiles of the dSphs are well known
demanding challenges for CDM models. Other Local group information is
also important: what is the stellar Initial Mass Function, what is the
distribution of chemical elements, what is the age range in the
Galactic Bulge, when did the last significant disk merger happen...?
All these challenge our appreciation of galaxy formation and
evolution, and in turn provide the information needed to refine the
models.  The partnership between local observations and {\sl ab initio}
theory is close, and developing well, and must culminate in GAIA.

There are clear similarities and distinctions between fundamental
properties of the different Galactic stellar populations, such as age,
metallicity, star formation history, angular momentum (figure~1) which
allow study of their individual histories. This is arguably one of the
greatest advantages of studies of Local Group galaxies: one is able to
disentangle the many different histories which have led to a galaxy
typical of those which dominate the luminosity density of the
Universe.

\section{GAIA and Dark Matter Mapping}

Ninety percent of the matter in the Universe is of an unknown nature
(to say nothing here of the even larger amount of dark energy). This
matter dominates gravitational potential wells in the early Universe
on all scales, and everywhere except in the centres of large galaxies
and dense stellar clusters today. The smallest scale lengths on which
this dark matter is dominant are an important constraint on its
nature: for example, if dark matter is concentrated on very small
scale lengths it cannot be relativistic. The only known way in which
we can study dark matter is through its dynamical effects on test
particles orbiting in a potential which it generates. The simplest
dark-matter dominated systems known, with scale lengths of interest to
constrain the physical nature of the dark matter, and with a dynamical
structure simple enough to be amenable to understanding, are the
Galactic satellite dwarf spheroidal galaxies. Very extensive studies,
using gravitational lensing, galaxy rotation curves, stellar and gas
kinematics, X-ray profiles, and so on, have initiated mapping dark
matter on larger scales. For the dwarf galaxies, stellar kinematics
are uniquely the method of choice.

This dark matter dominates the matter in the Universe, but remains
weakly constrained: what is the nature of the dark matter? A variety
of theoretical studies suggest that the small scale structure of dark
matter is the key test of its nature (Navarro, Frenk \& White 1997;
Klypin etal 1999; Moore etal 1999). Many extant studies of dark matter
on many different scale lengths are available, using techniques such
as gravitational lensing, X-ray luminosity and temperature profiles,
HI and H$\alpha$ rotation curves, and stellar dynamics. None has been
able to provide a reliable study of dark matter on the crucial small
length scales, because the only small dark-matter dominated relatively
simple dynamical systems, the Milky Way satellite dwarf galaxies, can
be studied in detail only with very accurate kinematics of faint
stars.

\subsection{The Importance of the Smallest Dark Haloes}

To be confined on small scales, a system must be cold.  For
self-gravitating systems the virial velocity, and equivalently virial
temperature, of a system in equilibrium is simply related to its mean
density, $\rho$, and characteristic radius, $R$, as
$$ V_{virial}^2 = k_B T_{virial}/m_p \sim {G \rho R^2}.$$

Baryonic material, which can radiate away energy while gaseous, can
form systems of a wide range of virial velocities, with the lower
limit just set by the cooling law.  In particular, baryonic dark
matter could form dark haloes of small scalelength.  One can relate
the `temperature' of non-baryonic dark matter (assumed in thermal
equilibrium in the early Universe) with its velocity dispersion at the
epoch in the early universe when it decouples from the ordinary matter
(Bond \& Szalay 1983).  `Hot' dark matter, for example a massive
neutrino, which is relativistic at this epoch, has a large
free-streaming length and cannot be confined on small scales (Bond,
Szalay \& White 1983).  In contrast, `cold' dark matter, which is
non-relativistic when it decouples from ordinary matter, can form
small-scale structure. One currently popular non-baryonic cold dark
matter candidate, the axion, is not produced through processes in
thermal equilibrium, has an extremely low `temperature', and  can
cluster on scales below that of even a dwarf galaxy.

Thus the study of dark matter in systems with the smallest
characteristic radius provides a thermometer of dark matter and can be
used to determine its nature. Dwarf spheroidal galaxies have (stellar)
scale-lengths of $\sim 300$~pc (Irwin \& Hatzidimitriou 1995),
comparable to the scale height of the Galactic thin disk and bulge.  
Numerical simulations of cosmological structure
formation do not have the dynamic range necessary to study such
scales.  Irrespective of the future developments of computing power
and techniques, the simulation of a dwarf galaxy requires the
inclusion of the unknown physics of star formation and feedback,
expected to be particularly important in these low velocity-dispersion
systems (e.g. Dekel \& Silk 1986; Wyse \& Silk 1985; Lin \& Murray
1991).  Improved understanding must be led by observations.

What are the smallest systems for which there is evidence of dark matter? 
Galactic disks are thin, of small scale-height, and
we are located in the middle of one.  Dark matter confined to the disk
(as distinct from the extended, Galactic-scale dark halo) would be
`cold'. However, the local neighbourhood is not a good place to study
cold disk dark matter, since it is now accepted there is no evidence
that any exists (Kuijken \& Gilmore 1989, 1991; Gilmore, Wyse \&
Kuijken 1989; Flynn \& Fuchs 1994;
Creze etal.\ 1998; Holmberg \& Flynn 2000).  Globular clusters are
fairly nearby, and have very small scalelengths, tens of parsecs, but
again there is no evidence for dark matter associated with globular
clusters (e.g. Meylan \& Heggie 1997).  Note that in fact in these two
cases there is not just lack of evidence for cold dark matter, there
is actual evidence for a lack of dark matter, which in itself is
significant.

In contrast, the dark matter content of low-luminosity dwarf galaxies,
as inferred from analyses of their internal stellar and gas
kinematics, makes them the most dark-matter-dominated galaxies
(e.g. Mateo 1998; Carignan \& Beaulieu 1989). Of these small galaxies,
the low-luminosity gas-free dwarf Spheroidals (dSph) are the most extreme.
Available stellar kinematic studies provide strong
evidence for the presence of dominant dark matter (e.g. review of
Mateo 1998), confirming earlier inferences from estimates of the
dSph's tidal radii and a model of the Milky Way gravitational
potential (Faber \& Lin 1983). 

\subsubsection{What information is essential?}

Substantial analytic methodologies have been developed over the last
few years by several groups to interpret colour-magnitude data and
stellar kinematics. These studies provide an existence theorem proving
that efficient analysis of the GAIA and complementary data is possible.

Adequate kinematic data  can determine the
potential well locally on a variety of different scales, parsecs up to
1Mpc.  The motivation is that the {\it smallest\/} scale is set by a
fundamental characteristic of the dark matter, its velocity
dispersion, analogously its temperature, so that study of the dynamics
on these scales provides unique and strong constraints on the nature
of dark matter.  Such a determination is possible
-- and is possible only -- using full 6-dimensional phase space data,
such as will be determinable by GAIA.

\section{Dwarf galaxies: an example analysis}

The dynamical structure of a dwarf galaxy, while simple, is not
trivial. The galaxies are probably triaxial, can sustain complex
families of anisotropic orbits, and live in a time-varying Galactic
tidal field. Given this complexity, considerable information is
required to derive a robust measurement of the 3-dimensional
gravitational potential, and hence the generating (dark) mass. Several
groups have
devised methods to explore the possible dynamical structure of a
Galactic satellite, and used these to simulate observational
approaches to determination of the mass distribution. These studies
show that determination of at least five
phase space coordinates -- two projected spatial coordinates, and all
three components of the velocity -- are necessary
and sufficient to provide a robust and reliable determination of a
dark matter mass distribution. Acquisition of these data are possible
only with a combination of  radial velocity data and
proper motion data for a well-selected sample of stars distributed
across the whole face of the target galaxy. In the nearest appropriate
target galaxies, Draco and Ursa Minor, the stars of interest are at
magnitude V$\sim 19$, and the required proper motion accuracy,
corresponding to 1-2 km/s in velocity, is 3--6$\mu as$yr$^{-1}$.  Thus
this specific test of the nature of matter on small scales is
better with SIM rather than GAIA. Nonetheless, the analysis
principles apply in general, and can be generalised to GAIA studies of
the local galaxy, where SIM cannot contribute.

An important simplification in any  dynamical analysis is that
the potential be time-independent over an internal dynamical timescale, 
the tracer
stellar distribution be in equilibrium with the potential, and be well-mixed
(relaxed). While dSph galaxies show every possible star formation
history, it is a defining characteristic that star formation ceased at
least a few Gyr ago (e.g. Hernandez, Gilmore \& Valls-Gabaud
2000). The internal dynamical times of these galaxies are typically
only $ t_{dyn} \sim R/\sigma \sim 2 \times 10^7 (R/200{\rm pc})(10
{\rm kms}^{-1}/\sigma)$~yr; thus star formation indeed ceased many
hundreds of crossing times ago and the systems should be well-mixed.
This assumption can of course be directly tested by the data.

\subsubsection{The Degeneracy between Anisotropy and Mass}

\begin{figure}[h!t]
%{\epsscale{0.4}
\plotfiddle{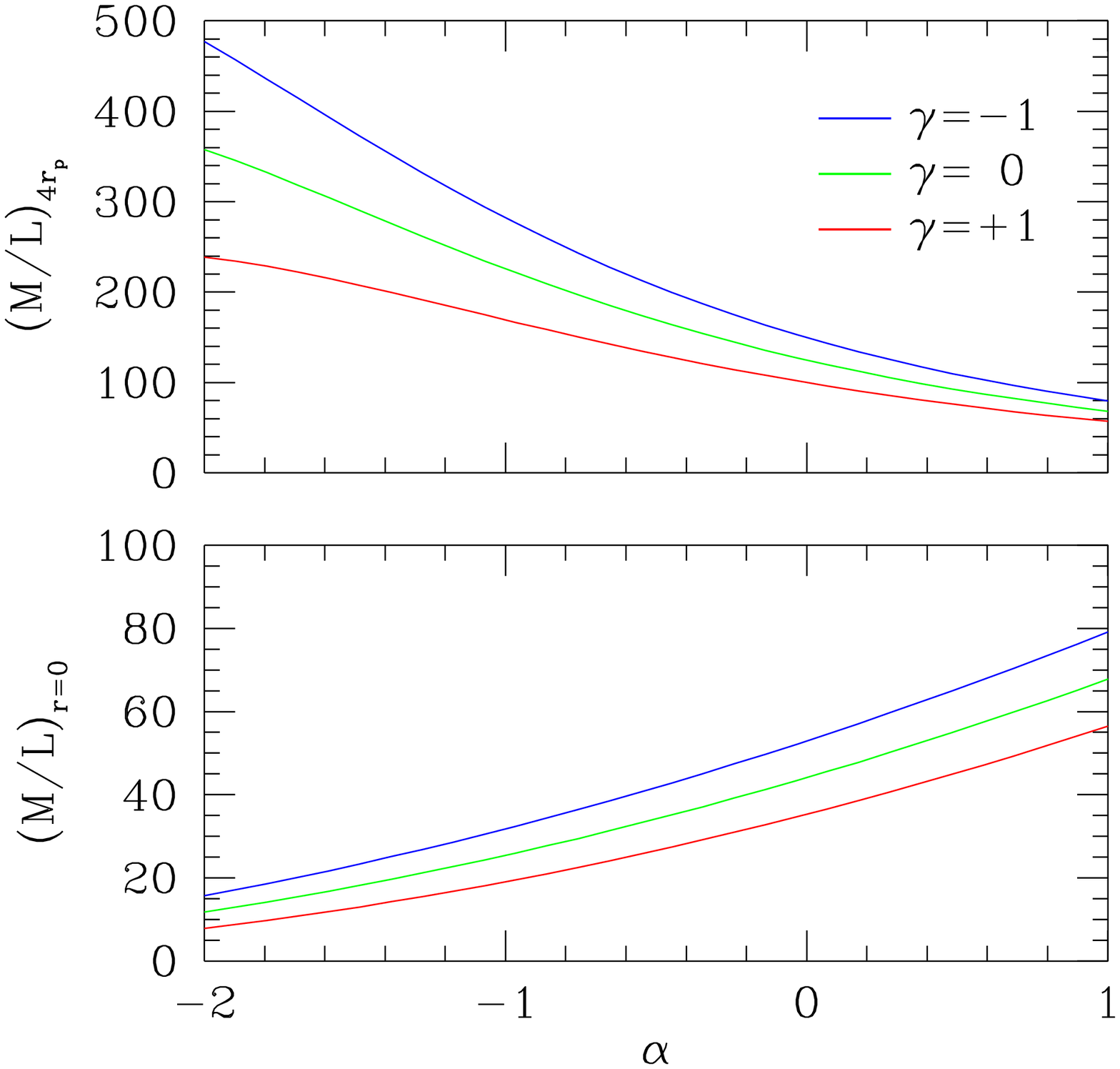}{7cm}{0}{40}{40}{-150}{-75}
\caption{\label{mlqbeta_fig} Variation of mass--to--light ratio (in
units of $M_\odot/L_\odot$) as a function of $\alpha$ for three
different $\gamma$ values.  The top panel shows the total $M/L$ within
$4 r_0$ ($3 r_0 \approx$ Draco King tidal radius; Irwin \&
Hatzidimitrou, 1995).  The bottom panel shows the central $M/L$.}
\end{figure}

Most studies of galactic dynamics have been impeded by the degeneracy
between anisotropy and mass.  This degeneracy, that one cannot distinguish
between gravitational potential gradients and gradients in orbital
anisotropy from measurements of just one component of the velocity
ellipsoid (radial velocity), is the fundamental limitation in extant
analyses, and the key justification for extending observation to 5
phase space coordinates. The large observed central radial
velocity dispersion is compatible with either a massive halo, and a
low central density; or no halo, and a large central density.  To
demonstrate our ability to discern between these possibilities,
Wilkinson etal (2002)
constructed a set of models that span the space of halo mass and
anisotropy, and perform Monte Carlo recoveries of the input anisotropy
and halo mass.

\begin{figure}
 \plottwo{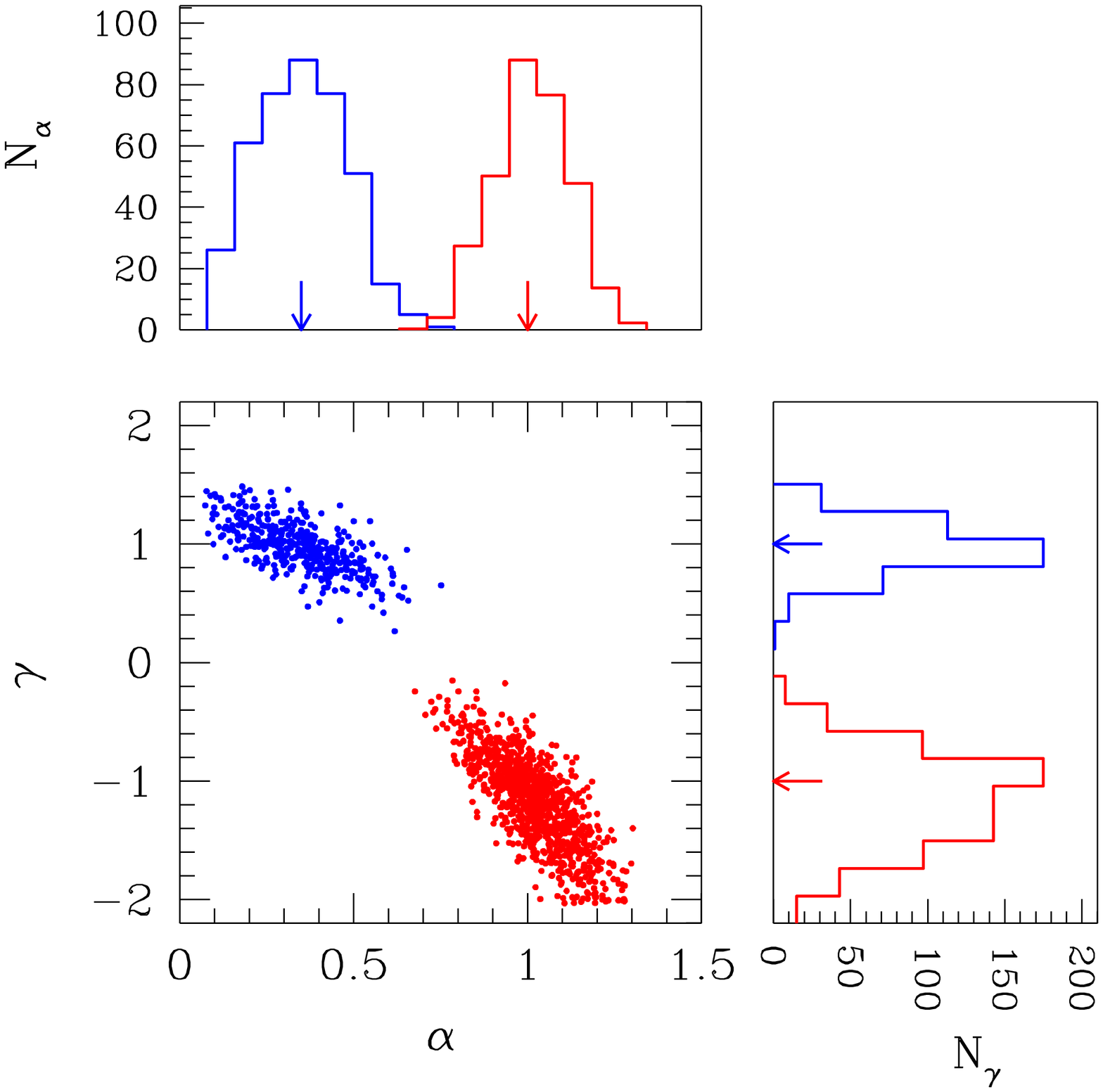}{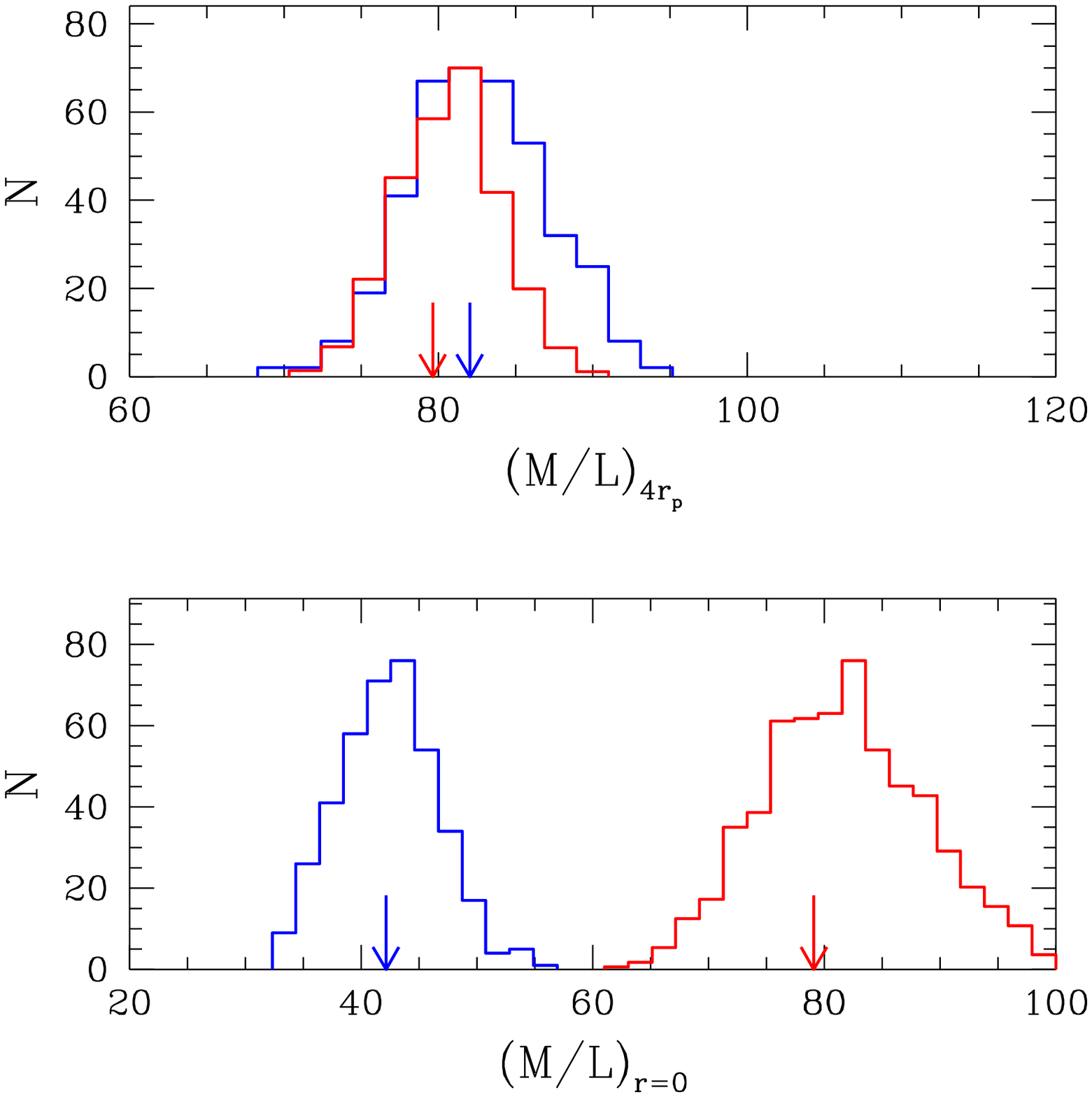}
%\epsscale{0.75}
%\plotone{alpha_beta_fig.ps}
\caption{\label{alpha_q_recovery_fig} {\sl Left:} Recovery of
best--fit $\alpha,\gamma$ for 2 sets of 250 star
artificial data ensembles; one set of ensembles  was created
with $\alpha=0.2, \gamma=1$, the other  with $\alpha=1,
\gamma=-1$.  The centre panel is a scatter plot of the recovered
$\alpha,\gamma$; the top and right panels are histograms of $\alpha$
and $\gamma$, respectively; arrows indicate actual values.  We are
able to distinguish clearly between these models, and break the
mass/anisotropy degeneracy. {\sl Right:} Recovered mass to light ratio
for simulations in left plot.  Top panel shows the total $M/L$ (solar
units) within 4 Plummer radii; bottom panel shows the central $M/L$.}
\end{figure}

They assume that the luminosity density of a dSph is given by a
Plummer model
\begin{equation}
\label{plummer_rho_eq}
\rho_{\rm p}(r)={\rho_0 \over \left(1+\left(r/r_0\right)^2\right)^{5/2}}\, ,
\end{equation}
where $\rho_0$ is determined by the total observed
luminosity.  Next, assume that the potential of the system has the
form
\begin{equation}
\label{model_psi_eq}
\psi(r)={\psi_0 \over \left(1+\left(r/r_0\right)^2\right)^{\alpha/2}}
\end{equation}

For this dark matter potential, $\alpha=1$ corresponds to a
mass--follows--light Plummer potential, Keplerian at large radii;
$\alpha=0$ yields, for large $r$, a flat rotation curve; and
$\alpha=-2$ gives a harmonic oscilator potential. As the parameter
$\alpha$ decreases, the dSph becomes more and more dark matter
dominated.

Finally, assume that the distribution function (DF) of the stars in
the model potential is a function entirely of the energy $E$ and norm
of the angular momentum $L$.  Under this assumption, one can derive a two
parameter set of DFs, parameterised by $\alpha$ and an anisotropy
parameter $\gamma$, namely
\begin{eqnarray}
F(E,L^2) = {\rho_0\over \psi_0^{5/\beta -\gamma/\beta}}
           {\Gamma (5/\beta - \gamma/\beta +1) \over
           (2\pi)^{3/2} \Gamma (5/\beta - \gamma/\beta - 1/2)}
           E^{5/\beta - \gamma/\beta - 3/2}\nonumber \\
           {}_2F_1(\gamma/2;3/2 - 5/\beta + \gamma/\beta,1, L^2/2E )
\end{eqnarray}

This formula holds for $L^2 < 2E$ and $\alpha <0$; similar expressions
can be derived for all the other cases.  These DFs generalise earlier
calculations by Dejonghe (1987), which are restricted to the case
where there is no dark matter.  In terms of Binney's anisotropy
parameter $\beta$, the radial and tangential velocity dispersions
$\sigma_r^2$ and $\sigma_\theta^2$ vary as
\begin{equation}
  \beta = 1 - {\sigma_\theta^2 \over \sigma_r^2} = {\gamma\over 2}
  {r^2\over 1+r^2}
\end{equation}

\begin{figure}[h!]
\plotfiddle{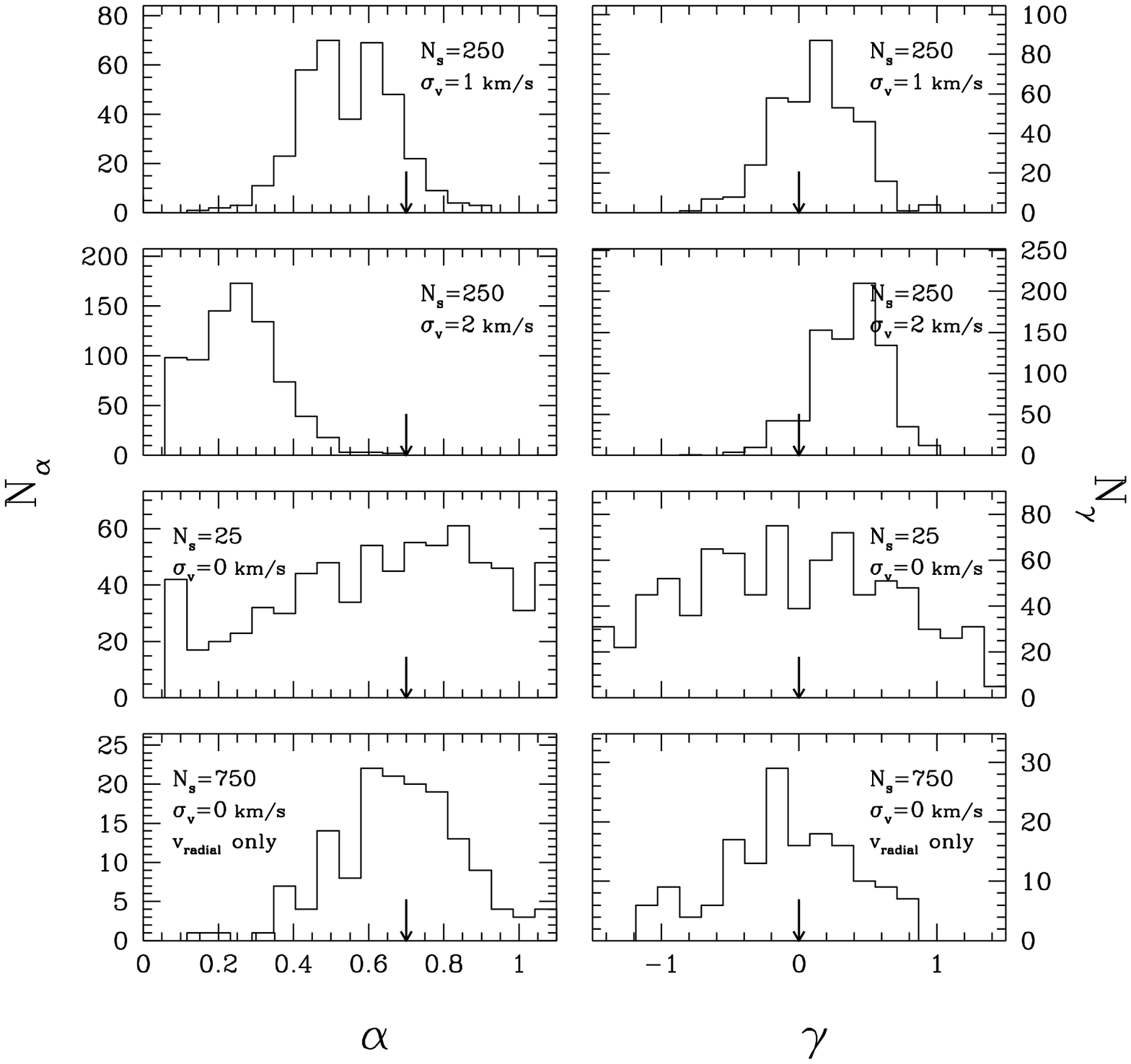}{8cm}{0}{50}{50}{-180}{-100}
\caption{ {\sl Top three panels: } Effect of velocity measurement
errors of 10\%, 20\%, and 40\% of the system dispersion (=10km/s
here), respectively, on recovery of halo parameter $\alpha$ (left) and
anisotropy parameter $\gamma$ (right).  The input (arrows) $\alpha$
and $\gamma$ can be recovered with measuring errors of 10\% or 20\% of
the system internal velocity dispersion, albeit with a (calibratable)
systematic shift in $\gamma$ (see text).  {\sl Third panel from top:}
Recovery of $\alpha$ and $\gamma$ using only 25 velocities; no
meaningful constraints can be placed on $\alpha$ and $\gamma$.  {\sl
Bottom panel:} Recovery of $\alpha$ and $\gamma $ using 750 radial
velocities; $\alpha$ and $\gamma$ are poorly constrained,
demonstrating the necessity of proper motions and radial velocity
data.  }\label{bqerr_fig}
\end{figure}

The potential normalisation $\psi_0$ in eqn~2 is a
well defined function of the observed central radial velocity
dispersion, and the assumed $\alpha$ and $\gamma$.  Figure~2
shows the variation of the central and large--scale
mass to light ratio as a function of $\alpha$ and $\gamma$.

To simulate our ability to resolve among values of $\gamma$ and
$\alpha$, we generate 6D phase space coordinates
$\{x_i,y_i,{v_x}_i,{v_y}_i,{v_z}_i\}_{i=1\ldots N}$ drawn from the DF 
$F(x,y,z,v_x,v_y,v_z;\alpha,\gamma)$, discard the $z$ (geocentric
radial position) coordinate, and attempt to recover $\gamma$ and
$\alpha$ using a Bayesian likelihood technique (see e.g. Little \&
Tremaine 1987, Kochanek 1996, Wilkinson \& Evans 1999, van der Marel
etal 2000).  Specifically,
we scan a grid of $\alpha,\gamma$, and at each point compute the
probability of observing the input data set
\begin{equation}
\label{Prob_eq}
P( \{x_i,y_i,{v_x}_i,{v_y}_i,{v_z}_i\}_{i=1\ldots N} | \alpha,\gamma)
= \prod_{i=1}^N \int_{-\infty}^\infty 
dz\, F(x_i,y_i,z,{v_x}_i,{v_y}_i, {v_z}_i ; \alpha,\gamma)
\end{equation}

By Bayes's theorem, and the assumption of uniform prior probabilities
of $\alpha,\gamma$, the most likely $\alpha,\gamma$ are given by
maximising eqn~5.  Confidence regions are obtained by
applying 2D $\chi^2$ statistics to the logarithm of
eqn~5.

\subsubsection{Results}
Figure~3 (left) shows the result of a Monte
Carlo simulation to recover the input $\alpha$ and $\beta$ of two
model distributions ($\alpha=0.35, \gamma=1$, and $\alpha=0.8, \gamma=-1$,
respectively).  These values of $\alpha$ and $\gamma$ lie along the
direction of the usual mass/anisotropy indeterminacy, and thus illustrate
our ability to break this degeneracy.  Both model fits involve 800
reconstructions, each of which contains an ensemble of 250 simulated
stars.
The right panel of Figure~3 depicts
constraints placed on the central $M/L$, and the total $M/L$
within four Plummer radii.  Although the two models evaluated
have a similar total mass within $4 r_p$, our method clearly
recovers the significant difference in central $M/L$ caused
by a halo. 

The top two panels of Figure~4 illustrate the effect of
anticipated velocity measurement errors which are 10\% and 20\% of the
internal (local) velocity dispersion of the (sub-)system under study,
assuming a 250 star sample.  These errors cause a systematic
displacement of the best--fit $\alpha$ in the small halo
($\alpha=0.7$) case we examine, because they create a significant
increase of the apparent velocity dispersion at the maximum radius ($3
r_P$) considered.  However, the width of the distribution remains
narrow, and the systematic $\alpha$ shift will be rigorously
incorporated into the analysis by convolving the velocity errors with
the DF.  Larger errors would not be acceptable, however, because
resolving among $\alpha$ values requires a precise determination of the
falloff of the velocity dispersion. 
The third panel of  Figure~4 illustrates the
recovery of $\alpha,\gamma$ using only 25 velocities; 
the halo and anisotropy are essentially unconstrained by
this small sample.  
The bottom panel  shows the effect of using 750 purely radial
velocities, with zero errors.  Although we have the
same number of one--dimensional velocities as in the
250 star cases, we are unable to recover $\alpha$ and $\gamma$;
this failure indicates that the analysis requires all three
velocity components, and this work could not be performed
using ground--based radial velocities.

\subsection{Triaxial dSph Models}

A triaxial spheroid seen projected on the plane of sky mimics a
rotationally symmetric ellipse. Triaxial systems can support much
more complex orbital structures than can 2-D systems; this allows a
degeneracy between radially-dependent orbital complexity and system
shape, potentially invalidating dynamical analyses. More complex
modelling is required in this case, and has been partially developed.
The (very obvious!) conclusion from extensive numerical simulations
based on those triaxial models is that it is vital to have full
kinematics available if we are to investigate triaxial (more
generally: not isolated spherical) systems.
 
\section{Implications for GAIA design}

The discussion above illustrates the considerable progress in
quantitative analyses of stellar kinematics which are currently under
development. the introduction recalled that GAIA is funded to study
Galactic evolution. What data must GAIA provide so these analysis
techniques can deliver this science. The quantitative analysis
techniques need distances, 2-D (better 3-D) kinematics, and some
astrophysical information on the distribution function of stellar
properties. Highly detailed information on a few stars is not what is
needed. Can GAIA deliver? Simulations presented at this meeting show
GAIA radial velocity data will be unable to provide useful scientific
data at low Galactic latitudes. Low Galactic latitudes, while ``only''
a few percent of sky, contain most of the stars, and contain ALL of the
inner disk and bulge. It is worth emphasising that these parts of the
Milky way are the only parts of high redshift galaxies which we can
see. If GAIA fails to provide adequate data on these low latitudes, it
fails a primary science goal.

However, as the simulations show, bad data are useless. The lesson for
GAIA spectra is clear: do something excellently, not several things
badly, optimise for the faintest possible stars in
uncrowded regions: leave low latitudes for the other instruments.

What other instruments? The medium band photometr (MBP) unfortunately
has been degraded to poor ground-based quality spatial resolution,
~1arcsec. Thus, as long experience has establsihed, MBP will also be
of little or no value in the inner Galaxy. Again, the implication is
clear: optimise MBP for relatively low extinction higher-latitude
regions, with somewhat metal-poor old stars. 

Can GAIA meets its primary science goals at all: only if the broad
band filters are optimised to deliver the essential minimum
astrophysical data. That is, the limited sensitivity of RVS and the
poor resolution of MBP require that the broad band filters be
optimised for astrophysical analyses, and not other
considerations. Without these choices, GAAI will fail to meet its
design science goals. With them, it will revolutionise our
understanding of galaxy formation and evolution.

\end{document}